\documentclass[prl,superscriptaddress,twocolumn]{revtex4}
\usepackage{enumerate}
\usepackage{amsfonts,amssymb,amsmath}
\usepackage[]{graphics,graphicx,epsfig}
\usepackage{amsthm}
\usepackage{float}

\usepackage{pdfpages}
\usepackage{epstopdf}
\DeclareGraphicsExtensions{.png,.pdf}
\usepackage{graphicx}
\usepackage{dcolumn}
\usepackage{natbib}
\usepackage{color}
\usepackage{multirow}
\usepackage{ulem}
\newcommand{\ket}[1]{|{#1}\rangle}

\usepackage{array}
\newcolumntype{P}[1]{>{\centering\arraybackslash}p{#1}}

\begin{document}

\title{Simple explanation of apparent Bell nonlocality of unentangled photons}
\author{Antoni W{\'o}jcik}
\affiliation{Institute of Spintronics and Quantum Information, Faculty of Physics, Adam Mickiewicz University, 61-614 Pozna\'n, Poland}

\author{Jan W{\'o}jcik}
\affiliation{Institute of Theoretical Physics and Astrophysics, University of Gdańsk, 80-308 Gda\'nsk, Poland}
\date{\today}

\begin{abstract}
Wang et al. [Science Advances, 1 Aug 2025 Vol 11, Issue 31] recently reported an experiment that they interpret as demonstrating a violation of Bell’s inequality using unentangled photons. Such a claim is highly controversial, since it is well established that product states cannot surpass the bounds set by local hidden variable models. In this manuscript, we analyze the essential features of the experiment through simplified, idealized scenarios. Our analysis shows that the apparent violation of Bell’s inequality originates from two key elements: postselection and an unconventional normalization procedure. These steps produce quantities that formally mimic Bell-type correlations but lack the operational meaning required in a genuine Bell test. We therefore argue that the reported violation does not demonstrate nonlocality without entanglement, but rather reflects a misinterpretation of otherwise valid and interesting multiphoton interference effects.
\end{abstract}

\maketitle

\section{Introduction}
One of the most remarkable features of the quantum realm, first highlighted by Bell \cite{Bell1,Bell2}, is that quantum-mechanical correlations between measurements performed in spatially separated laboratories can exceed the bounds imposed by local hidden variable (LHV) models. Subsequent experiments \cite{Clauser1,Aspect1,Aspect2,Aspect3,Zeilinger1} ruled out LHV models and confirmed the predictions of quantum mechanics through violations of Bell inequalities. The significance of this discovery was recognized with the 2022 Nobel Prize awarded to A. Aspect, J. Clauser, and A. Zeilinger.

A crucial prerequisite for such violations is that the particles measured in separate laboratories (commonly referred to as Alice’s and Bob’s) must be in an entangled state. Unentangled (product) states cannot surpass the bounds imposed by LHV models. It is therefore controversial that Wang et al. \cite{Wang1} presented results under the title “Violation of Bell inequality with unentangled photons.”. Unsurprisingly, such a claim sparked significant debate \cite{cieslinski1,wharton1}.

In this work, we seek to contribute to that discussion. Our aim is not to provide a detailed theoretical analysis of the experiment in question. Rather, we present a straightforward argument supporting the view that the apparent violation of the Bell inequality arises from a misinterpretation of otherwise highly important and unquestionably valid experimental results. To keep the main idea in focus, we avoid considering too many details and begin with an analysis of idealized, simple experiments (see Fig.1).

\section{Typical two-photon Bell scenario}
Let us start with the standard (idealized) two-photon CHSH (ref) scenario (see Fig.1a). Perfect source generates photon pairs in a maximally entangled state  
\begin{equation}
\frac{1}{\sqrt{2}}(\ket{HH}+\ket{VV}),
\end{equation}
and shares it between Alice and Bob. The quantum state is subsequently analyzed in Alice’s and Bob’s laboratories by means of polarizing beam splitters (PBS), followed by ideal single-photon detectors. In each laboratory, the PBS together with the two detectors implements a dichotomic projective measurement. The measurement settings are specified by the parameters $\alpha$ on Alice’s side and $\beta$ on Bob’s side. Accordingly, the measurement observable associated with Alice’s apparatus is
\begin{equation}
M_A(\alpha) = \cos(\alpha),\sigma_z + \sin(\alpha),\sigma_x,
\end{equation}
whereas the corresponding observable on Bob’s side is
\begin{equation}
M_B(\beta) = \cos(\beta),\sigma_z - \sin(\beta),\sigma_x,
\end{equation}
with $\sigma_x$ and $\sigma_z$ denoting the Pauli operators.

Dependence of coincidence rates on parameters $\alpha$ and $\beta$ is given by 
\begin{equation}
N(++|\alpha,\beta)=N_0 \ \cos^2\left(\frac{\alpha+\beta}{2}\right),
\end{equation}
\begin{equation}
N(-+|\alpha,\beta)=N_0 \ \sin^2\left(\frac{\alpha+\beta}{2}\right),
\end{equation}
\begin{equation}
N(+-|\alpha,\beta)=N_0 \ \sin^2\left(\frac{\alpha+\beta}{2}\right),
\end{equation}
\begin{equation}
N(--|\alpha,\beta)=N_0 \ \cos^2\left(\frac{\alpha+\beta}{2}\right),
\end{equation}
with $N_0$ dependent on the source intensity and independent from $\alpha$ and $\beta$.
These rates can be normalized with
\begin{equation}
p(jk|\alpha,\beta)=\frac{N(jk|\alpha,\beta)}{n(\alpha,\beta)},
\end{equation}
where $j,k=+,-$ and
\begin{gather}\nonumber
    n(\alpha,\beta) = N(++|\alpha,\beta)+N(+-|\alpha,\beta)+\\+N(-+|\alpha,\beta)+N(--|\alpha,\beta),
\end{gather}
leading to probabilities
\begin{equation}\label{p1}
p(++|\alpha,\beta)=\frac{1}{4}(1+\cos(\alpha+\beta)),
\end{equation}
\begin{equation}\label{p2}
p(-+|\alpha,\beta)=\frac{1}{4}(1-\cos(\alpha+\beta)),
\end{equation}
\begin{equation}\label{p3}
p(+-|\alpha,\beta)=\frac{1}{4}(1-\cos(\alpha+\beta)),
\end{equation}
\begin{equation}\label{p4}
p(--|\alpha,\beta)=\frac{1}{4}(1+\cos(\alpha+\beta)).
\end{equation}
With these probabilities, correlation functions can be defined as
\begin{gather}\nonumber
E(\alpha,\beta)=p(++|\alpha,\beta)-p(-+|\alpha,\beta)-\\\label{e2}-p(+-|\alpha,\beta)+p(--|\alpha,\beta),
\end{gather}
which in this case simplifies to 
\begin{equation}\label{e}
E(\alpha,\beta)=\cos(\alpha+\beta).
\end{equation}
Finally, the Bell parameter 
\begin{gather}\nonumber
B(\alpha_1,\alpha_2,\beta_1,\beta_2)=|-E(\alpha_1,\beta_1)+\\\label{b}+E(\alpha_1,\beta_2)+E(\alpha_2,\beta_1)+E(\alpha_2,\beta_2)|.
\end{gather}
can approach Tsirelson bound $2\sqrt{2}$ violating Bell inequality $B\leq2$ (e.g. for $(\alpha_1=\pi/2,\alpha_2=0,\beta_1=\pi/4,\beta_2=-\pi/4)$)

\begin{figure}[t]
\includegraphics[width=0.4\textwidth]{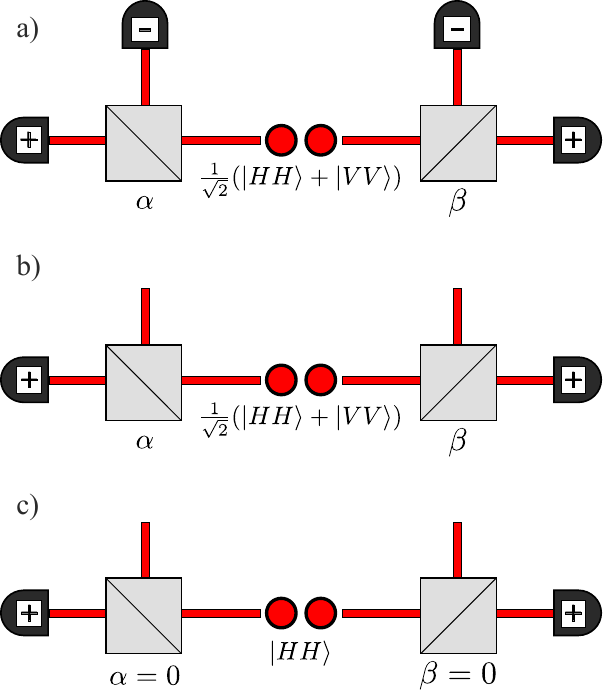}
\caption{Three experimental scenarios: a) the standard two-photon Bell test, in which a pair of entangled photons is distributed to spatially separated laboratories and their polarizations are measured using PBS and four single-photon detectors;
b) a modified two-photon Bell test employing only two detectors; c) a scenario where two unentangled photons are distributed to spatially separated laboratories, yielding an apparent violation of a Bell inequality.}
\label{f1}
\end{figure}
\section{Two-photon Bell scenario with only two detectors}
Imagine now that we have only two detectors (let us say both denoted by $+$) (see Fig.1b). There is no way to directly measure coincidence rates $N(-+|\alpha,\beta)$, $N(+-|\alpha,\beta)$ and $N(--|\alpha,\beta)$. However, we can make use of Eqs. \ref{p1} - \ref{p4} and observe that
\begin{equation}
N(-+|\alpha,\beta)=N(++|\alpha+\pi,\beta),
\end{equation}
\begin{equation}
N(+-|\alpha,\beta)=N(++|\alpha,\beta+\pi),
\end{equation}
\begin{equation}
N(--|\alpha,\beta)=N(++|\alpha+\pi,\beta+\pi).
\end{equation}
This allows us to propose another definition of probabilities 
\begin{equation}\label{t1}
\tilde{p}(++|\alpha,\beta)=\frac{N(++|\alpha,\beta)}{\tilde{n}(\alpha_,\beta)},
\end{equation}
\begin{equation}\label{t2}
\tilde{p}(-+|\alpha,\beta)=\frac{N(++|\alpha+\pi,\beta)}{\tilde{n}(\alpha_,\beta)},
\end{equation}
\begin{equation}\label{t3}
\tilde{p}(+-|\alpha,\beta)=\frac{N(++|\alpha,\beta+\pi)}{\tilde{n}(\alpha_,\beta)},
\end{equation}
\begin{equation}\label{t4}
\tilde{p}(--|\alpha,\beta)=\frac{N(++|\alpha+\pi,\beta+\pi)}{\tilde{n}(\alpha_,\beta)},
\end{equation}

where the normalization constant is given by
\begin{gather}\nonumber
\tilde{n}(\alpha_,\beta)=N(++|\alpha,\beta)+N(++|\alpha+\pi,\beta)+\\\label{t5}+N(++|\alpha,\beta+\pi)+N(++|\alpha+\pi,\beta+\pi),
\end{gather}

To avoid confusion, we will denote these new probabilities with a tilde. 
Note that in this kind of experiment $\tilde{p}(++|\alpha,\beta)=p(++|\alpha,\beta)$, so Bell inequality violation can also be claimed in such a way.

\section{Apparent Bell inequality violation}
The third experiment, which we propose to analyze, will lead to apparent Bell inequality violation. We again use only two detectors (see Fig.1c). We fix PBS at $\alpha=\beta=0$ and use the photon source controlled by two parameters $\alpha'$ and $\beta'$. The source emits photon pairs in the product state 
\begin{equation}
\ket{HH},
\end{equation}
not with a constant rate, but proportional to  
\begin{equation}
N_0=\cos^2\left(\frac{\alpha'+\beta'}{2}\right).
\end{equation}
We can now measure the coincidence rate $N(++|\alpha',\beta')$ and use it to define probabilities dependent on $\alpha'$ and $\beta'$. With proper normalization (for $\alpha'+\beta'\neq \pi$)
\begin{equation}\label{p1p}
p(++|\alpha',\beta')=\frac{N(++|\alpha',\beta')}{N_0}=1,
\end{equation}
Let us recall that we consider ideal sources, PBS and detectors.
Consequently, it must be that
\begin{equation}\label{p2}
p(-+|\alpha',\beta')=0,
\end{equation}
\begin{equation}
p(+-|\alpha',\beta')=0,
\end{equation}
\begin{equation}
p(--|\alpha',\beta')=0.
\end{equation}
Thus $E(\alpha',\beta')=1$ is independent on $\alpha'$ and $\beta'$. Bell parameter must be $B=2$ as it should be with the photons in the product state.

However, if one insists on defining four functions of $\alpha'$ and $\beta'$ in analogy with Eqs. \ref{t1} - \ref{t5} the result is
\begin{equation}\label{q1}
q(++|\alpha',\beta')=\frac{1}{4}(1+\cos(\alpha'+\beta')),
\end{equation}
\begin{equation}\label{q2}
q(-+|\alpha',\beta')=\frac{1}{4}(1-\cos(\alpha'+\beta')),
\end{equation}
\begin{equation}\label{q3}
q(+-|\alpha',\beta')=\frac{1}{4}(1-\cos(\alpha'+\beta')),
\end{equation}
\begin{equation}\label{q4}
q(--|\alpha',\beta')=\frac{1}{4}(1+\cos(\alpha'+\beta')).
\end{equation}
We denote the above functions $q$ to further highlight the difference between them and functions given by \ref{t1} - \ref{t4}. Although formally
\begin{equation}
q(jk|\alpha',\beta')=\tilde{p}(jk|\alpha,\beta),
\end{equation}
their interpretation is quite different. Functions $q(jk|\alpha',\beta')$ can hardly be interpreted as probabilities because there are no events that correspond to them. Of course, if one use $q(jk|\alpha',\beta')$ instead of $p(jk|\alpha,\beta)$ in Eqs. \ref{e2}, \ref{b} it will be possible to obtain $B=2 \sqrt{2}$. However, this has nothing to do with non-locality and LHV models. This is because parameters $\alpha'$ and $\beta'$ do not correspond to local measurement settings as they should for deriving the Bell inequality.

\section{Four-photon Bell experiment}
Let us now return to the original paper of Wang et al. \cite{Wang1}. They observed a fascinating quantum-mechanical effect based on multiphoton interference and indistinguishability. However, when interpreted as a violation of Bell’s inequality, their arguments are not convincing. Despite the experimental complexity, the essence of the effect observed corresponds to our apparent Bell inequality violation scenario. Two aspects are crucial here: postselection and normalization.

The detailed form of the state generated in the experiment is provided in the work by Cieśliński et al. \cite{cieslinski1}. The postselection procedure accepts only those events in which a single photon is simultaneously detected in each of the four detectors used in the experiment. With such postselection, the source effectively generates an unentangled state
\begin{equation}
\ket{VHHV},
\end{equation}
with a rate proportional to
\begin{equation}
\cos^2\left(\frac{\alpha+\beta}{2}\right).
\end{equation}
Thus, as a result of postselection, the parameters $\alpha$ and $\beta$ should not be regarded as parameters of local measurements, but rather as variables controlling the effective source intensity (as in our simplified example). Moreover, the measured coincidence rate $N(++|\alpha,\beta)$ is not properly normalized, but is used to compute "probabilities" in the same manner as we defined by $q(jk|\alpha',\beta')$. This further strengthens the correspondence between our apparent Bell inequality violation thought experiment and the experiment under discussion. On this basis, we claim that it is the postselection and normalization that account for the observed apparent Bell inequality violation.

\section{Conclusions}
The experiment of Wang et al. \cite{Wang1} reveals an elegant quantum interference phenomenon based on multiphoton indistinguishability. However, when interpreted as a violation of Bell inequality with unentangled photons, the conclusions are misleading. By analyzing simplified, idealized scenarios, we demonstrated that the apparent violation arises from two key features: postselection and normalization of coincidence rates. These procedures transform experimental data into quantities that mimic Bell-type correlations but lack the operational meaning required in a genuine Bell test.

Our analysis highlights that the parameters $\alpha$ and $\beta$ in the Wang et al. experiment do not correspond to local measurement settings, as assumed in the derivation of Bell inequalities, but instead control the effective source intensity. Consequently, the reported Bell parameter exceeding the LHV bound does not signify nonlocal correlations.

The results of Wang et al. remain important, as they explore novel aspects of multiphoton interference, yet they should not be interpreted as evidence of Bell nonlocality without entanglement. We hope that our analysis clarifies this distinction and contributes to a more precise understanding of the role of entanglement in foundational quantum experiments.
\section{Acknowledgements}
AW was supported by the National Science Centre (NCN, Poland) under the Maestro Grant no. DEC-2019/34/A/ST2/00081. JW was supported by the National Science Centre (NCN, Poland) within the Chist-ERA project (Grant No. 2023/05/Y/ST2/00005).

\bibliographystyle{apsrev}

\end{document}